\documentclass[lettersize,journal]{IEEEtran}

\usepackage{amsmath,amssymb,amsfonts,amsthm}
\usepackage{graphicx}
\usepackage{textcomp}
\usepackage{stfloats}
\usepackage{url}
\usepackage{verbatim}
\usepackage{cite}
\usepackage{subfig}
\usepackage{booktabs}
\usepackage{multirow}
\usepackage{array}
\usepackage[ruled,vlined]{algorithm2e}

\begin{document}


\title{Joint Pinching-Antenna Position and Power Optimization for Dual-Waveguide ISAC Systems



}


\author{
Saeid Pakravan,
Imene Trigui,
Wessam Ajib,
and Wei-Ping Zhu
\thanks{
S. Pakravan and W. Ajib are with the Department of Computer Science, University of Quebec in Montreal (UQAM), Montreal, QC, Canada
(e-mail: pakravan.saeid@uqam.ca; ajib.wessam@uqam.ca).
}
\thanks{
I. Trigui is with the Department of Applied Sciences, University of Quebec at Chicoutimi (UQAC), Chicoutimi, QC, Canada
(e-mail: itrigui@uqac.ca).
}
\thanks{
W.-P. Zhu is with the Department of Electrical and Computer Engineering, Concordia University, Montreal, QC, Canada
(e-mail: weiping@ece.concordia.ca).
}
}

\maketitle


\begin{abstract}

Pinching-antenna (PA) systems provide spatial reconfigurability for integrated sensing and communication (ISAC) through controllable radiation locations along dielectric waveguides. This paper investigates a dual-waveguide PA-assisted ISAC system in which the communication user also serves as the sensing target, coupling the communication and sensing links through a common user/target geometry. Spatially separated transmit and receive waveguides provide separate spatial control over the forward and return sensing paths. We jointly optimize the transmit and receive PA positions and transmit power to maximize the aggregate communication rate subject to minimum sensing-SNR and two-dimensional localization Cramér–Rao bound (CRB) requirements, along with power, energy, and deployment constraints. To solve the resulting non-convex problem, we develop an alternating optimization algorithm that combines a bounded water-filling power update with a proximal first-order joint PA-position update. Numerical results show that the proposed design achieves higher communication rates than the considered benchmarks, with more pronounced gains under stringent localization requirements.

\end{abstract}




\section{Introduction}

Integrated sensing and communication (ISAC) has emerged as a key enabling paradigm for future wireless networks by allowing communication and sensing functionalities to share spectrum, hardware, and signaling resources \cite{10608156}. Compared with conventional architectures in which communication and sensing are designed separately, ISAC can improve spectral efficiency, reduce hardware overhead, and enable environment-aware wireless services. These capabilities have motivated growing interest in ISAC for applications such as intelligent transportation, smart manufacturing, and location-aware wireless services \cite{10944644,11404239}.

Despite its significant potential, the performance of ISAC systems is fundamentally constrained by the underlying antenna architecture. Most existing ISAC designs rely on fixed-position antenna arrays, whose spatial configuration remains unchanged and whose communication and sensing responses are primarily adapted through beamforming. This fixed geometry limits the ability to reconfigure propagation paths and sensing geometries in response to changing user locations, blockage, and environmental conditions. These limitations have motivated the exploration of antenna architectures in which the spatial configuration itself becomes an additional design dimension. Such spatial reconfigurability is particularly relevant to ISAC, since changes in the antenna geometry can affect not only communication channels but also the propagation and observation geometries that govern sensing performance.

In this context, pinching antenna (PA) systems have recently emerged as a promising waveguide-based architecture for spatially reconfigurable wireless transmission \cite{10945421, 11625963, 11364174, 11481959, pakravan2026pass}. In PA systems, electromagnetic signals propagate through dielectric waveguides and are coupled into free space through controllable pinching locations. By adjusting these radiation locations along the waveguide, PA systems introduce additional spatial degrees of freedom for modifying propagation distances and channel phase responses \cite{11457366}. Unlike conventional antenna arrays with fixed radiation points, PA systems enable flexible spatial configuration through guided signal propagation over dielectric waveguides \cite{11434944}. These characteristics make PA systems particularly attractive for ISAC, where spatial reconfigurability can be exploited not only to enhance communication links but also to adapt sensing geometries and echo propagation for joint communication and sensing optimization.

\subsection{Related Works}

Existing studies on PA-assisted ISAC span performance characterization, resource optimization, multi-waveguide transceiver design, and localization-oriented sensing \cite{11414134,11122551,11212813,11197530,11303725,11586289,11597999,11111111,11424157,11587609,11586698,4444444}. The communication--sensing trade-off enabled by configurable PA positions was investigated in \cite{11414134}, where the achievable communication--sensing rate region was characterized and PA-position optimization was shown to enlarge the achievable region relative to conventional fixed-antenna systems. Beyond performance characterization, resource-allocation approaches have been developed to exploit PA reconfigurability under joint communication and sensing requirements. In \cite{11122551}, the PA positions and transmit power were jointly optimized to maximize the communication rate subject to sensing and energy constraints using a maximum-entropy-based reinforcement-learning approach.

Multi-waveguide and separated transmit--receive PA architectures have also been investigated for ISAC. In \cite{11212813}, a multi-waveguide architecture coordinating transmit and receive PAs was considered, in which the transmit PA positions and beamforming were jointly designed to maximize the communication rate subject to a radar signal-to-noise ratio (SNR) requirement. A separated two-waveguide architecture was investigated in \cite{11197530}, where one waveguide was used to transmit the joint communication--sensing signal while the other collected the reflected echoes, and the illumination power was maximized subject to communication quality-of-service (QoS) requirements. Distributed multi-target PA-assisted ISAC was further studied in \cite{11303725}, where transmit and receive PA positions and receive beamforming were jointly optimized to improve sensing fairness and mitigate inter-target interference. Low-complexity algorithms for adaptive transmit/receive PA-position tuning were developed in \cite{11586289}. More recently, a sensing-centric PA-assisted ISAC system with separated transmit and receive waveguides was investigated in \cite{11597999}, where power allocation, transmit/receive pinching locations, signaling, and carrier frequency were jointly designed.

Beyond conventional sensing-performance metrics, estimation-theoretic measures have increasingly been incorporated into PA-assisted ISAC design. The Cramér–Rao bound (CRB) for PA-assisted localization was analyzed in \cite{11111111}, showing that PA reconfigurability can improve positioning accuracy. In \cite{11424157}, a PA system transmitting and uniform linear array (ULA)-receiving architecture was considered, where the sensing CRB was minimized subject to communication-QoS, power-budget, and PA-deployment constraints. Reflection-aware PA-assisted sensing was investigated in \cite{11587609}, where the asymmetric propagation of the transmit and reflected sensing paths was explicitly modeled and incorporated into CRB-oriented PA design. Angle-estimation CRB minimization through joint pinching and baseband beamforming was further investigated for multi-waveguide PA-assisted ISAC systems in \cite{11586698}. More recently, localization CRB minimization with joint PA-position and beamforming design was studied for segmented-waveguide PA architectures in \cite{4444444}.

Table~\ref{tab:comparison} summarizes representative PA-assisted ISAC studies in terms of their localization objectives, transmit/receive architectures, and optimization variables.

\begin{table*}[t]
\centering
\caption{Comparison of Representative PA-Assisted ISAC Studies}
\label{tab:comparison}

\renewcommand{\arraystretch}{1.25}
\setlength{\tabcolsep}{3pt}
\scriptsize

\begin{tabular}{
p{0.8cm}
c
c
c
c
p{10cm}
}
\toprule
\textbf{Ref.} &
\shortstack{\textbf{Position}\\\textbf{Localization}} &
\shortstack{\textbf{Separate}\\\textbf{Tx/Rx}} &
\shortstack{\textbf{Joint Tx/Rx}\\\textbf{PA Positioning}} &
\shortstack{\textbf{Power}\\\textbf{Allocation}} &
\multicolumn{1}{c}{\textbf{Main Contribution}}
\\
\midrule

\cite{11414134}
& $\times$ & $\times$ & $\times$ & $\times$
& Characterized the communication--sensing rate region via PA-position optimization.
\\

\cite{11122551}
& $\times$ & $\times$ & $\times$ & $\checkmark$
& Developed learning-based PA-position and power optimization under sensing and energy constraints.
\\

\cite{11212813}
& $\times$ & $\checkmark$ & $\times$ & $\times$
& Investigated multi-waveguide ISAC with transmit-PA positioning and beamforming.
\\

\cite{11197530}
& $\times$ & $\checkmark$ & $\checkmark$ & $\checkmark$
& Investigated separated transmit/receive waveguides for communication and echo sensing.
\\

\cite{11303725}
& $\times$ & $\checkmark$ & $\checkmark$ & $\times$
& Developed multi-target ISAC with joint transmit/receive PA positioning and receive beamforming.
\\

\cite{11586289}
& $\times$ & $\checkmark$ & $\checkmark$ & $\checkmark$
& Developed low-complexity transmit/receive PA tuning for communication--sensing enhancement.
\\

\cite{11597999}
& $\times$ & $\checkmark$ & $\checkmark$ & $\checkmark$
& Jointly designed transmit/receive PA locations, power, signaling, and carrier frequency.
\\

\cite{11111111}
& $\checkmark$ & $\times$ & $\times$ & $\times$
& Analyzed localization CRB and the positioning benefits of PA reconfigurability.
\\

\cite{11424157}
& $\checkmark$ & $\times$ & $\times$ & $\times$
& Minimized sensing CRB with PA transmission and ULA reception.
\\

\cite{11587609}
& $\times$ & $\times$ & $\times$ & $\checkmark$
& Developed reflection-aware CRB-oriented PA-position and power optimization.
\\

\cite{11586698}
& $\checkmark$ & $\checkmark$ & $\times$ & $\times$
& Minimized angle-estimation CRB via joint pinching and baseband beamforming.
\\

\cite{4444444}
& $\checkmark$ & $\checkmark$ & $\checkmark$ & $\times$
& Minimized localization CRB via joint PA-position and beamforming design.
\\

\midrule

\textbf{This work}
& $\checkmark$ & $\checkmark$ & $\checkmark$ & $\checkmark$
& \textbf{Jointly optimizes transmit/receive PA positions and power under sensing-SNR and localization-CRB constraints.}
\\

\bottomrule
\end{tabular}
\end{table*}

\subsection{Motivation and Main Contributions}

Despite the recent progress in PA-assisted ISAC, several challenges remain in localization-oriented system design. First, while sensing metrics such as sensing SNR, sensing rate, and illumination power have been widely considered, recent studies have also incorporated CRB-based localization metrics. The integration of explicit localization-accuracy requirements into communication-oriented PA design remains less explored, especially when the communication user itself serves as the localization target. Second, although separated transmit and receive PA architectures have been investigated, jointly configuring the transmit and receive PA positions for localization-aware ISAC introduces coupled sensing geometries that must be coordinated with communication performance. Third, jointly optimizing transmit and receive PA positions and transmit power under simultaneous communication, sensing-SNR, and localization-CRB requirements leads to a highly coupled non-convex problem due to the interplay among waveguide propagation, sensing geometry, and communication performance.

Motivated by these considerations, this paper investigates a localization-oriented dual-waveguide PA-assisted ISAC architecture in which the communication user simultaneously serves as the sensing target. Consequently, the user's location affects both the communication channel and the sensing geometry, coupling communication and localization through a common spatial parameter. The transmit and receive PAs are deployed on spatially separated waveguides, allowing their locations to be configured according to both communication and localization requirements. Based on unified communication and echo-based sensing models, we jointly optimize the transmit/receive PA positions and transmit power to balance communication performance and localization requirements.

The main contributions of this paper are summarized as follows:

\begin{itemize}

\item We develop a localization-oriented dual-waveguide PA-assisted ISAC framework in which the communication user simultaneously serves as the sensing target. The spatially separated waveguides enable separate configuration of the transmit and receive PA positions according to communication and localization requirements.

\item We establish communication and echo-based sensing models that account for guided-wave propagation, free-space transmission, and the cascaded transmit–target–receive response. Based on the sensing model, sensing quality and localization accuracy are characterized through the sensing SNR and position CRB, respectively.

\item We formulate a joint transmit/receive PA-position and transmit-power optimization problem that maximizes the achievable communication rate subject to sensing-SNR, localization-CRB, energy-budget, antenna-spacing, and waveguide-deployment constraints.

\item We develop an alternating optimization (AO) framework to address the resulting non-convex problem. The transmit-power subproblem admits a water-filling-type solution under the sensing and localization requirements, while the PA positions are iteratively optimized subject to the coupled spatial and deployment constraints.

\item Numerical results demonstrate communication-rate gains over the considered fixed-position, partially optimized, and single-waveguide benchmarks, particularly under stringent localization requirements.

\end{itemize}

The remainder of this paper is organized as follows. Section~II presents the dual-waveguide PA-assisted ISAC system model, develops the communication and localization-oriented sensing models, and formulates the joint PA-position and transmit-power optimization problem. Section~III develops the proposed AO-based solution. Section~IV presents the numerical results, and Section~V concludes the paper.

\textbf{Notation:} Throughout this paper, boldface lowercase and uppercase letters denote vectors and matrices, respectively. $(\cdot)^T$ and $(\cdot)^*$ denote transpose and complex conjugate operations, respectively. $\|\cdot\|$ and $|\cdot|$ denote the Euclidean norm and the magnitude of a scalar, respectively. $\Re\{\cdot\}$ denotes the real part of a complex quantity. $\mathcal{CN}(0,\sigma^2)$ represents a circularly symmetric complex Gaussian distribution with zero mean and variance $\sigma^2$. $\mathrm{tr}(\cdot)$ and $(\cdot)^{-1}$ denote the trace and matrix inverse, respectively, and $\mathbf{A}\succ0$ indicates that $\mathbf{A}$ is positive definite. The operator $\nabla$ denotes the gradient with respect to the indicated variables.

\section{System Model and Problem Formulation}
\label{sec:system_model}

\begin{figure}[t]
    \centering
    \includegraphics[width=7.5cm,height=4.5cm]{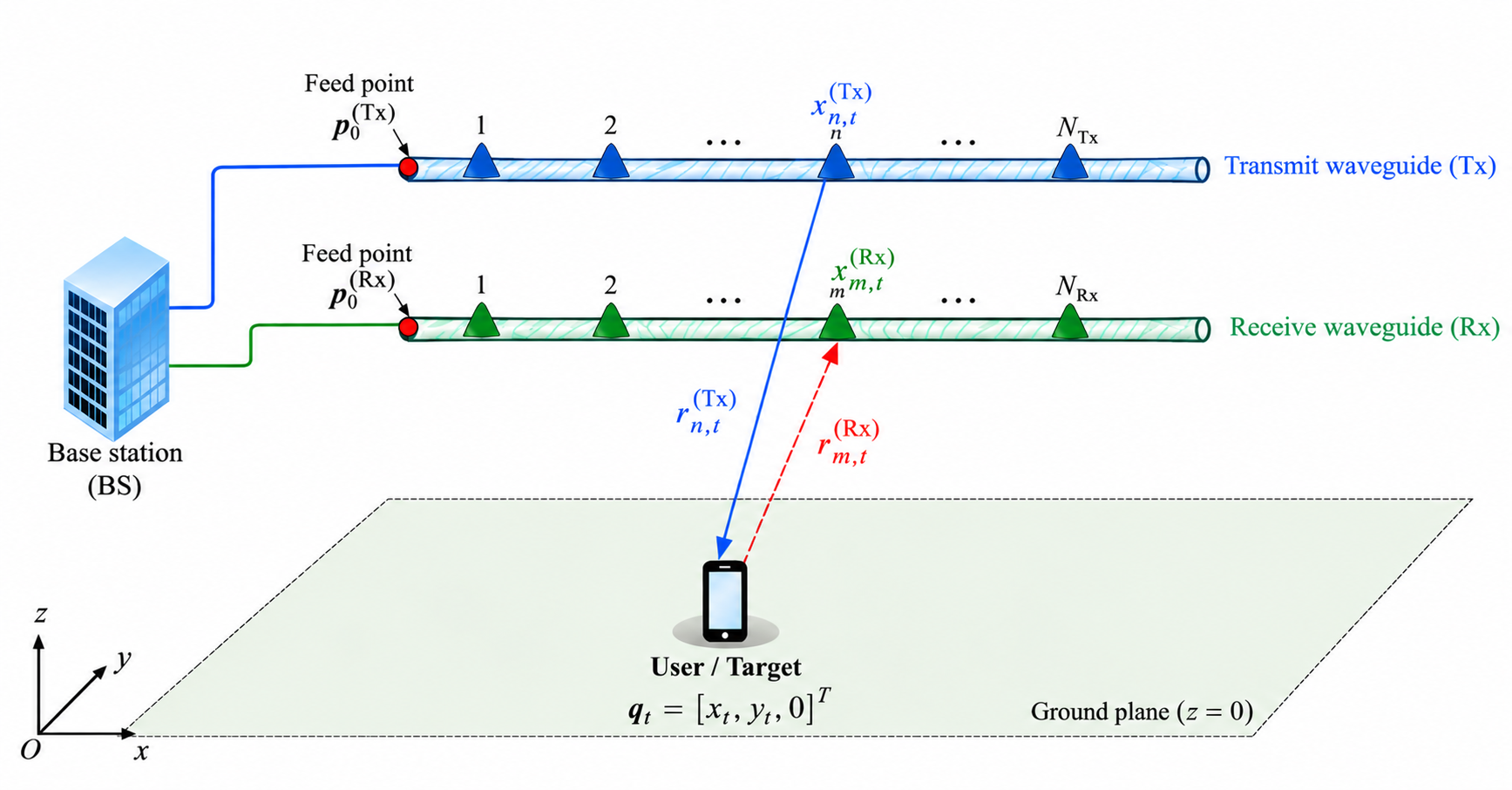}
    \caption{Dual-waveguide PA-assisted ISAC architecture, where the communication user simultaneously serves as the sensing target.}
    \label{fig:system_model}
\end{figure}

We consider a PA-assisted ISAC system in which a base station (BS) communicates with a wireless user that simultaneously serves as the sensing target. The system operates over time slots indexed by
$t\in\mathcal{T}\triangleq\{1,\ldots,T\}$.
The BS employs two spatially separated dielectric waveguides: a transmit waveguide equipped with transmit PAs (TPAs) and a receive waveguide equipped with receive PAs (RPAs). The transmit waveguide radiates the dual-functional communication--sensing signal, whereas the receive waveguide collects the signal reflected by the user/target. Fig.~\ref{fig:system_model} illustrates the considered dual-waveguide PA-assisted ISAC architecture.

\subsection{System Geometry and PA Deployment}

We adopt a three-dimensional Cartesian coordinate system. In time slot $t$, the user/target is located on the ground plane at $\mathbf{q}_t=[x_t,y_t,0]^T.
    \label{eq:user_position}$
Let $i\in\{\mathrm{Tx},\mathrm{Rx}\}$ index the transmit and receive waveguides. The set of PAs deployed on waveguide $i$ is denoted by
$\mathcal{N}_i\triangleq\{1,\ldots,N_i\}$,
where $N_{\mathrm{Tx}}$ and $N_{\mathrm{Rx}}$ denote the numbers of TPAs and RPAs, respectively.
Waveguide $i$ extends along the $x$-axis over a length $L_i$. The $k$-th PA on waveguide $i$ in time slot $t$ is located at
$
\mathbf{p}_{k,t}^{(i)}
=
[x_{k,t}^{(i)},y_i,d_i]^T,
k\in\mathcal{N}_i,
$
where $y_i$ and $d_i$ specify the fixed lateral position and height of waveguide $i$, respectively, while $x_{k,t}^{(i)}\in[-L_i/2,L_i/2]$ denotes the configurable PA position along the waveguide.
The PA positions are assumed to be reconfigurable at the beginning of each time slot and remain fixed within the slot.
The two waveguides are spatially separated such that
$
    (y_{\mathrm{Tx}},d_{\mathrm{Tx}})
    \neq
    (y_{\mathrm{Rx}},d_{\mathrm{Rx}}).
$
The feed point of waveguide $i$ is located at the beginning of the waveguide and is given by
$
\mathbf{p}_{0}^{(i)}
=
[-L_i/2,y_i,d_i]^T.
$

The free-space propagation distance between the user/target and the $k$-th PA on waveguide $i$ is
\begin{equation}
    r_{k,t}^{(i)}
    =
    \left\|
    \mathbf{q}_t-\mathbf{p}_{k,t}^{(i)}
    \right\|
    =
    \sqrt{
    (x_t-x_{k,t}^{(i)})^2
    +(y_t-y_i)^2
    +d_i^2
    }.
    \label{eq:free_space_distance}
\end{equation}
   
Accordingly, the guided propagation distance from the feed point to the $k$-th PA is
\begin{equation}
\ell_{k,t}^{(i)}
=
x_{k,t}^{(i)}
+
\frac{L_i}{2}.
\label{eq:waveguide_distance}
\end{equation}
For compactness, the PA-position vector on waveguide $i$ is defined as
$\mathbf{x}_{t}^{(i)}
    =[x_{1,t}^{(i)},\ldots,x_{N_i,t}^{(i)}]^T.
    \label{eq:pa_location_vector}$

\subsection{Waveguide-Aware Channel Model}

Let $\lambda$ denote the free-space carrier wavelength and
$\lambda_g=\lambda/n_{\mathrm{eff}}$ denote the guided wavelength inside the dielectric waveguide, where $n_{\mathrm{eff}}$ is the effective refractive index. We adopt a line-of-sight (LoS)-dominated spherical-wave propagation model that accounts for both free-space propagation and guided-wave propagation along the dielectric waveguides \cite{11626907, 11595855, 11589280}. Small-scale fading and mutual coupling among neighboring PAs are neglected.

Let $\rho_k^{(i)}$ denote the fixed coupling coefficient associated with the $k$-th PA on waveguide $i$. For the transmit waveguide, the coupling coefficients satisfy
$
    \sum_{k=1}^{N_{\mathrm{Tx}}}
    \rho_k^{(\mathrm{Tx})}
    =1,\
    \rho_k^{(\mathrm{Tx})}\geq 0,
    \label{eq:coupling_normalization}$
so that $P_t$ represents the transmit-power budget at the input of the transmit waveguide. Equal coupling,
$\rho_k^{(\mathrm{Tx})}=1/N_{\mathrm{Tx}}$,
is a special case. For the receive waveguide,
$\rho_k^{(\mathrm{Rx})}$ denotes the fixed effective collection/coupling efficiency of the $k$-th RPA and is not an optimization variable.

The contribution of the $k$-th PA on waveguide $i$ to the effective propagation response is modeled as
\begin{equation}
    a_{k,t}^{(i)}
    =
    \sqrt{\rho_k^{(i)}}
    \frac{\alpha}{r_{k,t}^{(i)}}
    \exp\left(
    -\kappa_i\ell_{k,t}^{(i)}
    \right)
    \exp\left(
    -j\phi_{k,t}^{(i)}
    \right),
    \label{eq:pa_channel}
\end{equation}
where
$\alpha=\lambda/(4\pi)$,
$\kappa_i$ denotes the amplitude attenuation coefficient of waveguide $i$, and
\begin{equation}
    \phi_{k,t}^{(i)}
    =
    \frac{2\pi}{\lambda}r_{k,t}^{(i)}
    +
    \frac{2\pi}{\lambda_g}\ell_{k,t}^{(i)}
    \label{eq:total_phase}
\end{equation}
accounts for the phase shifts accumulated during free-space and guided-wave propagation.

Accordingly, the effective propagation response associated with waveguide $i$ is
\begin{equation}
    g_t^{(i)}
    =
    g^{(i)}
    \left(
    \mathbf{x}_{t}^{(i)},\mathbf{q}_t
    \right)
    =
    \sum_{k=1}^{N_i}
    a_{k,t}^{(i)}.
    \label{eq:1unified_channel}
\end{equation}
The model in \eqref{eq:1unified_channel} is used for both the BS-to-user transmit response
$g_t^{(\mathrm{Tx})}$
and the user/target-to-BS receive response
$g_t^{(\mathrm{Rx})}$.
The functional dependence is omitted when no ambiguity arises.

\subsection{Communication and Echo-Sensing Models}

In time slot $t$, the BS transmits a normalized dual-functional waveform $u_t$ with unit average power, i.e., $\mathbb{E}[|u_t|^2]=1$. With total transmit power $P_t$, the received communication signal at the user is
\begin{equation}
    y_t^{\mathrm{c}}
    =
    \sqrt{P_t}\,
    g_t^{(\mathrm{Tx})}u_t
    +
    n_t^{\mathrm{c}},
    \label{eq:communication_signal}
\end{equation}
where
$n_t^{\mathrm{c}}\sim\mathcal{CN}(0,\sigma_c^2)$
denotes the communication receiver noise. The corresponding communication SNR is
\begin{equation}
    \gamma_t^{\mathrm{c}}
    =
    \frac{
    P_t|g_t^{(\mathrm{Tx})}|^2
    }{\sigma_c^2},
    \label{eq:communication_snr}
\end{equation}
and the achievable communication rate is
\begin{equation}
    R_t
    =
    B\log_2
    \left(
    1+
    \frac{
    P_t|g_t^{(\mathrm{Tx})}|^2
    }{\sigma_c^2}
    \right),
    \label{eq:rate}
\end{equation}
where $B$ denotes the system bandwidth.

Since the communication user simultaneously serves as the sensing target, the transmitted waveform is reflected by the same node and collected through the receive waveguide. The user/target is modeled as a point scatterer, and the sensing echo follows the cascaded transmit--target--receive spherical-wave response. Let $\beta_t$ denote the effective complex sensing coefficient of the user/target, incorporating its scattering response and other fixed sensing-link factors. We assume that $\beta_t$ is available from a preceding sensing/calibration stage and remains constant within each sensing interval. Accordingly, the localization CRB is evaluated conditionally on $\beta_t$. After coherent processing of the known transmitted waveform, the effective sensing observation in time slot \(t\) is modeled as
\begin{equation}
    y_t^{\mathrm{s}}
    =
    \sqrt{P_t}\,
    \beta_t
    g_t^{(\mathrm{Rx})}
    g_t^{(\mathrm{Tx})}
    u_t
    +
    n_t^{\mathrm{s}},
    \label{eq:echo_signal_dual}
\end{equation}
where
$n_t^{\mathrm{s}}\sim\mathcal{CN}(0,\sigma_s^2)$
denotes the effective noise after sensing processing. Since the transmitted dual-functional waveform is known at the BS, $u_t$ is treated as known in the sensing processing. The corresponding sensing SNR is
\begin{equation}
    \gamma_t^{\mathrm{s}}
    =
    \frac{
    P_t|\beta_t|^2
    |g_t^{(\mathrm{Tx})}|^2
    |g_t^{(\mathrm{Rx})}|^2
    }{\sigma_s^2}.
    \label{eq:sensing_snr_dual}
\end{equation}
The sensing SNR characterizes the quality of the received echo and is used to impose a minimum sensing-reliability requirement. The separate transmit and receive waveguides provide distinct configurable propagation responses for the forward and reflected sensing paths.

\subsection{Localization Model}

While the sensing SNR in \eqref{eq:sensing_snr_dual} characterizes the received echo quality, it does not directly quantify the accuracy with which the user/target position can be estimated. We therefore characterize localization performance through the CRB.

The unknown two-dimensional target-location parameter vector in time slot $t$ is defined as
$\boldsymbol{\xi}_t
    =
    [x_t,y_t]^T.
    \label{eq:target_parameter}$
At the beginning of each time slot, the BS is assumed to have a nominal location estimate
$
    \bar{\mathbf{q}}_t
    =
    [\bar{x}_t,\bar{y}_t,0]^T,$
which may be obtained from a preceding localization or tracking stage. The PA configuration is designed based on this nominal location, and the Fisher information matrix (FIM) and CRB are evaluated locally at
$\boldsymbol{\xi}_t=\bar{\boldsymbol{\xi}}_t$,
where
$\bar{\boldsymbol{\xi}}_t=[\bar{x}_t,\bar{y}_t]^T$.
For notational simplicity, the dependence on the nominal operating point is omitted hereafter.

From \eqref{eq:echo_signal_dual}, the noise-free sensing observation is
\begin{equation}
    \mu_t(\boldsymbol{\xi}_t)
    =
    \sqrt{P_t}\,
    \beta_t
    g_t^{(\mathrm{Rx})}
    g_t^{(\mathrm{Tx})}
    u_t.
    \label{eq:sensing_mean}
\end{equation}
For complex Gaussian sensing noise with parameter-independent variance $\sigma_s^2$, the $(a,b)$-th entry of the FIM for estimating $\boldsymbol{\xi}_t$ is
\begin{equation}
    [\mathbf{J}_t]_{a,b}
    =
    \frac{2}{\sigma_s^2}
    \Re
    \left\{
    \frac{\partial\mu_t^*}
    {\partial\xi_{a,t}}
    \frac{\partial\mu_t}
    {\partial\xi_{b,t}}
    \right\},
    \qquad
    a,b\in\{1,2\},
    \label{eq:fim_dual}
\end{equation}
where
$\xi_{1,t}=x_t$
and
$\xi_{2,t}=y_t$.

The localization CRB metric is defined as
\begin{equation}
    \mathrm{CRB}_t
    =
    \begin{cases}
    \mathrm{tr}\!\left(\mathbf{J}_t^{-1}\right),
    & \mathbf{J}_t \succ 0,\\[1mm]
    +\infty,
    & \text{otherwise},
    \end{cases}
    \label{eq:1crb_dual}
\end{equation}
which, for a nonsingular FIM, lower-bounds the sum of the estimation-error variances for the two location coordinates. The definition in \eqref{eq:1crb_dual} assigns an infinite localization-error bound to configurations that do not provide sufficient information for two-dimensional localization.

Applying the chain rule to \eqref{eq:sensing_mean} yields
\begin{equation}
    \frac{\partial\mu_t}
    {\partial\xi_{a,t}}
    =
    \sqrt{P_t}\,
    \beta_t u_t
    \left(
    g_t^{(\mathrm{Rx})}
    \frac{\partial g_t^{(\mathrm{Tx})}}
    {\partial\xi_{a,t}}
    +
    g_t^{(\mathrm{Tx})}
    \frac{\partial g_t^{(\mathrm{Rx})}}
    {\partial\xi_{a,t}}
    \right).
    \label{eq:mu_derivative}
\end{equation}

Since the guided propagation distance $\ell_{k,t}^{(i)}$ depends only on the PA position along waveguide $i$ and not on the target location, differentiating \eqref{eq:1unified_channel} with respect to $\xi_{a,t}$ gives
\begin{equation}
    \frac{\partial g_t^{(i)}}
    {\partial\xi_{a,t}}
    =
    \sum_{k=1}^{N_i}
    a_{k,t}^{(i)}
    \left(
    -\frac{1}{r_{k,t}^{(i)}}
    -
    j\frac{2\pi}{\lambda}
    \right)
    \frac{\partial r_{k,t}^{(i)}}
    {\partial\xi_{a,t}}.
    \label{eq:channel_derivative_location}
\end{equation}
From \eqref{eq:free_space_distance}, the required distance derivatives are
\begin{equation}
    \frac{\partial r_{k,t}^{(i)}}{\partial x_t}
    =
    \frac{
    x_t-x_{k,t}^{(i)}
    }{
    r_{k,t}^{(i)}
    },
    \qquad
    \frac{\partial r_{k,t}^{(i)}}{\partial y_t}
    =
    \frac{
    y_t-y_i
    }{
    r_{k,t}^{(i)}
    }.
    \label{eq:distance_derivatives}
\end{equation}
Equations
\eqref{eq:fim_dual}--\eqref{eq:distance_derivatives}
show that the transmit and receive PA positions jointly determine the spatial derivatives of the sensing response and, consequently, the localization information contained in the received echo.

\subsection{Problem Formulation}

We jointly optimize the transmit power and the transmit and receive PA positions to maximize the aggregate achievable communication rate over the considered time slots while satisfying sensing, localization, power, energy, and PA-deployment requirements.

The optimization variables are collected as
\begin{equation}
    \mathcal{X}
    =
    \left\{
    \mathbf{x}_{t}^{(\mathrm{Tx})},
    \mathbf{x}_{t}^{(\mathrm{Rx})},
    P_t
    \right\}_{t=1}^{T}.
    \label{eq:optimization_variables}
\end{equation}

The resulting optimization problem is formulated as
\begin{subequations}
\label{prob:main}
\begin{align}
    \max_{\mathcal{X}}
    \quad
    &
    \sum_{t=1}^{T}
    B\log_2
    \left(
    1+
    \frac{
    P_t|g_t^{(\mathrm{Tx})}|^2
    }{\sigma_c^2}
    \right)
    \label{prob:obj}
    \\
    \mathrm{s.t.}
    \quad
    &
    \frac{
    P_t|\beta_t|^2
    |g_t^{(\mathrm{Tx})}|^2
    |g_t^{(\mathrm{Rx})}|^2
    }{\sigma_s^2}
    \geq
    \Gamma_{\min},
    \quad
    \forall t\in\mathcal{T},
    \label{prob:snr}
    \\
    &
    \mathrm{CRB}_t
    \leq
    \varepsilon_{\max},
    \quad
    \forall t\in\mathcal{T},
    \label{prob:crb}
    \\
    &
    0\leq P_t\leq P_{\max},
    \quad
    \forall t\in\mathcal{T},
    \label{prob:power}
    \\
    &
    \sum_{t=1}^{T}
    P_t\Delta t
    \leq
    E_{\max},
    \label{prob:energy}
    \\
    &
    -\frac{L_i}{2}
    \leq
    x_{k,t}^{(i)}
    \leq
    \frac{L_i}{2},
    \quad
    \forall
    k\in\mathcal{N}_i,\,
    t\in\mathcal{T},\,
    i\in\{\mathrm{Tx},\mathrm{Rx}\},
    \label{prob:deployment}
    \\
    &
    x_{k+1,t}^{(i)}
    -
    x_{k,t}^{(i)}
    \geq
    \delta_i,
    \quad
    \substack{
    k=1,\ldots,N_i-1,\\
    t\in\mathcal{T},\,
    i\in\{\mathrm{Tx},\mathrm{Rx}\}.
    }
    \label{prob:spacing}
\end{align}
\end{subequations}

Constraint \eqref{prob:snr} imposes the minimum sensing-SNR requirement $\Gamma_{\min}$ to ensure sufficient echo quality, whereas \eqref{prob:crb} limits the localization CRB to the prescribed threshold $\varepsilon_{\max}$ to guarantee the required localization accuracy. Constraint \eqref{prob:power} imposes the per-slot transmit-power limit, and \eqref{prob:energy} limits the total transmission energy over the considered time horizon, where $\Delta t$ denotes the duration of each time slot. Constraint \eqref{prob:deployment} restricts each PA to its feasible deployment interval along the corresponding waveguide, while \eqref{prob:spacing} guarantees a minimum separation $\delta_i$ between adjacent PAs. The PAs on each waveguide are indexed according to their increasing positions along the waveguide.

Problem \eqref{prob:main} is non-convex due to the nonlinear and coupled dependence of the communication rate, sensing SNR, and localization CRB on the PA positions and transmit powers. In particular, the sensing-SNR constraint depends on the product of the transmit and receive effective channel gains, whereas the localization constraint involves the inverse of the PA-configuration-dependent FIM.

\section{Proposed AO-Based Solution}
\label{sec:solution}

Problem~\eqref{prob:main} is non-convex due to the coupled dependence of the communication rate, sensing SNR, and localization CRB on the PA positions and transmit powers. To obtain a tractable solution, we adopt an AO framework in which the transmit powers and PA positions are updated successively. For fixed PA positions, the transmit-power subproblem admits a closed-form water-filling-type solution. For fixed transmit powers, the transmit and receive PA positions are jointly refined using a proximal first-order approximation.

\subsection{Transmit-Power Optimization}
\label{subsec:power_optimization}

For fixed transmit and receive PA positions, the effective responses
$g_t^{(\mathrm{Tx})}$ and $g_t^{(\mathrm{Rx})}$ are constants. The transmit-power subproblem is therefore given by
\begin{subequations}
\label{prob:power_sub}
\begin{align}
    \max_{\{P_t\}}
    \quad
    &
    \sum_{t=1}^{T}
    B\log_2
    \left(
    1+
    \frac{
    P_t|g_t^{(\mathrm{Tx})}|^2
    }{\sigma_c^2}
    \right)
    \label{prob:power_sub_obj}
    \\
    \mathrm{s.t.}
    \quad
    &
    \eqref{prob:snr}-\eqref{prob:energy}.
\end{align}
\end{subequations}

From the sensing-SNR constraint in \eqref{prob:snr}, the minimum transmit power required in time slot $t$ is
\begin{equation}
    P_{t,\mathrm{SNR}}^{\min}
    =
    \frac{
    \Gamma_{\min}\sigma_s^2
    }{
    |\beta_t|^2
    |g_t^{(\mathrm{Tx})}|^2
    |g_t^{(\mathrm{Rx})}|^2
    }.
    \label{eq:min_snr_power}
\end{equation}

For fixed PA positions, the noise-free sensing response in
\eqref{eq:sensing_mean} is proportional to $\sqrt{P_t}$. Hence, the corresponding FIM scales linearly with $P_t$ as
$
    \mathbf{J}_t(P_t)
    =
    P_t\mathbf{J}_{t,0},
    \label{eq:fim_power_scaling}
$
where
$\mathbf{J}_{t,0}\triangleq\mathbf{J}_t(P_t=1)$
for the given PA configuration. If
$\mathbf{J}_{t,0}\succ0$, then
\begin{equation}
    \mathrm{CRB}_t
    =
    \frac{
    \mathrm{tr}
    \left(
    \mathbf{J}_{t,0}^{-1}
    \right)
    }{P_t}.
    \label{eq:crb_power_scaling}
\end{equation}
Define
$    c_t
    \triangleq
    \mathrm{tr}
    \left(
    \mathbf{J}_{t,0}^{-1}
    \right).
    \label{eq:ct_definition}
$
The CRB constraint in \eqref{prob:crb} then requires
\begin{equation}
    P_{t,\mathrm{CRB}}^{\min}
    =
    \frac{c_t}{\varepsilon_{\max}}.
    \label{eq:min_crb_power}
\end{equation}
Accordingly, the minimum feasible transmit power in time slot $t$ is
\begin{equation}
    P_t^{\min}
    =
    \max
    \left\{
    P_{t,\mathrm{SNR}}^{\min},
    P_{t,\mathrm{CRB}}^{\min}
    \right\}.
    \label{eq:minimum_feasible_power}
\end{equation}

For a given PA configuration, the power-allocation subproblem is feasible if and only if $\mathbf{J}_{t,0}\succ0,
    P_t^{\min}\leq P_{\max}$ for all $t\in\mathcal{T},
    \label{eq:power_feasibility_slot}$ and $
    \sum_{t=1}^{T}
    P_t^{\min}\Delta t
    \leq
    E_{\max}.
    \label{eq:power_feasibility_energy}
$
If $\mathbf{J}_{t,0}$ is singular for any time slot, increasing the transmit power cannot remove the corresponding localization degeneracy.

For a feasible PA configuration, Problem~\eqref{prob:power_sub} is a convex optimization problem. Define
\begin{equation}
    a_t
    \triangleq
    \frac{|g_t^{(\mathrm{Tx})}|^2}{\sigma_c^2}.
    \label{eq:at_definition}
\end{equation}
When the total-energy constraint is active, the Karush–Kuhn–Tucker (KKT) conditions yield
\begin{equation}
    P_t^\star
    =
    \min
    \left\{
    P_{\max},
    \max
    \left[
    P_t^{\min},
    \frac{B}
    {\nu\Delta t\ln 2}
    -
    \frac{1}{a_t}
    \right]
    \right\},
    \label{eq:waterfilling}
\end{equation}
where $\nu>0$ is the dual variable associated with the total-energy constraint and is selected such that
$
    \sum_{t=1}^{T}
    P_t^\star\Delta t
    =
    E_{\max}.
    \label{eq:energy_equality}
$
Since the allocated total energy is nonincreasing in $\nu$, the dual variable can be efficiently determined via bisection.
If
$
    \sum_{t=1}^{T}
    P_{\max}\Delta t
    \leq
    E_{\max},
    \label{eq:inactive_energy_condition}
$
the total-energy constraint is inactive. Since the communication objective is monotonically increasing in each $P_t$, the optimal solution then reduces to
$
    P_t^\star=P_{\max},
    \forall t\in\mathcal{T}.
    \label{eq:inactive_energy_solution}
$
A derivation of \eqref{eq:waterfilling} is provided in Appendix~A.

\subsection{Joint PA-Position Optimization}
\label{subsec:pa_optimization}

For fixed transmit powers, the transmit and receive PA positions are jointly optimized. This joint update is particularly relevant to the considered dual-waveguide architecture, since the receive-PA positions do not directly affect the communication rate but influence both the sensing SNR and the localization FIM. Consequently, optimizing the two PA sets jointly allows the receive-side spatial degrees of freedom to support the sensing and localization requirements while the transmit-PA positions are adjusted to improve the communication performance.

Define the joint PA-position vector at time slot $t$ as
\begin{equation}
\mathbf z_t
\triangleq
\begin{bmatrix}
(\mathbf x_t^{(\mathrm{Tx})})^T &
(\mathbf x_t^{(\mathrm{Rx})})^T
\end{bmatrix}^{T}.
\label{eq:joint_pa_vector}
\end{equation}
For notational convenience, let $\mathcal Z_t$ denote the feasible PA deployment set specified by the waveguide boundaries and minimum-spacing constraints in Problem~\eqref{prob:main}. 
With ${P_t}$ fixed, the PA-position subproblem can be expressed as
\begin{subequations}
\label{prob:pa_position}
\begin{align}
\max_{\{\mathbf z_t\}} \quad
& \sum_{t=1}^{T} R_t(\mathbf z_t)
\label{prob:pa_position_obj}\\
\mathrm{s.t.}\quad
& \gamma_t^{\mathrm{s}}(\mathbf z_t)
\geq \Gamma_{\min},
\quad \forall t\in\mathcal T,
\label{prob:pa_position_snr}\\
& \mathrm{CRB}_t(\mathbf z_t)
\leq \varepsilon_{\max},
\quad \forall t\in\mathcal T,
\label{prob:pa_position_crb}\\
& \mathbf z_t\in\mathcal Z_t,
\quad \forall t\in\mathcal T.
\label{prob:pa_position_set}
\end{align}
\end{subequations}
Problem~\eqref{prob:pa_position} remains non-convex because the free-space and guided-wave propagation terms depend nonlinearly on the PA positions. This dependence is further inherited by the sensing SNR and the FIM through the coherent transmit and receive channel responses.

To obtain a tractable position update, we construct a proximal first-order approximation around the current feasible configuration. Let $\mathbf z_t^{q}$ denote the joint PA-position vector at iteration $q$, and let $\mathbf d_t$ denote its displacement. For a differentiable scalar function $f_t(\mathbf z_t)$, define
\begin{equation}
\widetilde f_t^{q}(\mathbf d_t)
\triangleq
f_t(\mathbf z_t^{q})
+
\nabla_{\mathbf z_t}f_t(\mathbf z_t^{q})^{T}
\mathbf d_t.
\label{eq:joint_first_order_model}
\end{equation}
Accordingly, $\widetilde{\gamma}_t^{\mathrm{s},q}(\mathbf d_t)$ and
$\widetilde{\mathrm{CRB}}_t^{q}(\mathbf d_t)$ denote the local first-order models of the sensing SNR and CRB, respectively.

At iteration $q$, the joint PA search direction is obtained from
\begin{subequations}
\label{prob:joint_pa_update}
\begin{align}
\max_{\{\mathbf d_t\}} \quad
& \sum_{t=1}^{T}
\left[
\nabla_{\mathbf z_t} R_t(\mathbf z_t^{q})^{T}\mathbf d_t
-\frac{\|\mathbf d_t\|_2^2}{2\eta_q}
\right]
\label{prob:joint_pa_update_obj}\\
\mathrm{s.t.}\quad
& \widetilde{\gamma}_t^{\mathrm{s},q}(\mathbf d_t)
\geq \Gamma_{\min},
\quad \forall t\in\mathcal T,
\label{prob:joint_pa_update_snr}\\
& \widetilde{\mathrm{CRB}}_t^{q}(\mathbf d_t)
\leq \varepsilon_{\max},
\quad \forall t\in\mathcal T,
\label{prob:joint_pa_update_crb}\\
& \mathbf z_t^{q}+\mathbf d_t\in\mathcal Z_t,
\quad \forall t\in\mathcal T.
\label{prob:joint_pa_update_set}
\end{align}
\end{subequations}
where $\eta_q>0$ is the proximal parameter. The quadratic regularization term controls the displacement from the current PA configuration and renders the local objective strongly concave. Since the first-order sensing-SNR and CRB constraints, as well as the deployment and ordered minimum-spacing constraints, are affine in $\mathbf d_t$, Problem~\eqref{prob:joint_pa_update} is a convex quadratic program (QP). Moreover, since $\mathbf z_t^q$ is feasible for the original constraints, $\mathbf d_t=\mathbf 0$ is feasible for the local constraints, ensuring feasibility of the QP at each iteration. Because no inter-slot PA-motion constraint is imposed in Problem~\eqref{prob:main}, the local updates are separable across time slots and can be carried out in parallel.

Let $\widehat{\mathbf d}_t^{q}$ denote the solution of
\eqref{prob:joint_pa_update}. The PA positions are updated as
\begin{equation}
\mathbf z_t^{q+1}
=
\mathbf z_t^{q}
+
\alpha_q\widehat{\mathbf d}_t^{q},
\qquad
0<\alpha_q\leq1,
\label{eq:joint_pa_position_update}
\end{equation}
where $\alpha_q$ is determined by backtracking. Since the first-order constraint models are not necessarily inner approximations of the original nonlinear feasible set, each candidate configuration is evaluated using the original sensing-SNR and CRB expressions, and the step size is reduced whenever necessary to preserve feasibility. The accepted step is also required to yield a nondecreasing communication objective for the current fixed power allocation. If no acceptable step is found, the current PA configuration is retained.

The required gradients can be evaluated analytically from the position-dependent channel model. In particular, whenever $\mathbf J_t\succ 0$, the derivative of the localization CRB with respect to an arbitrary PA coordinate $z$ is
\begin{equation}
\frac{\partial\mathrm{CRB}_t}{\partial z}
=
-
\operatorname{tr}
\left(
\mathbf J_t^{-1}
\frac{\partial\mathbf J_t}{\partial z}
\mathbf J_t^{-1}
\right).
\label{eq:crb_position_gradient}
\end{equation}
The remaining channel, rate, sensing-SNR, and FIM derivatives are given in Appendix~B.

The proposed AO procedure is summarized in Algorithm~\ref{alg:AO}.

\RestyleAlgo{ruled}
\SetNlSty{textbf}{}{:}
\begin{algorithm}[t]
\SetAlgoLined

\textbf{Initialize:}
Choose an initial PA configuration
$\{\mathbf z_t^{0}\}_{t\in\mathcal T}$
that admits a feasible power allocation, convergence tolerance
$\epsilon$, and maximum number of iterations $I_{\max}$.
Set $q=0$.\\

Compute the initial power allocation
$\{P_t^{0}\}_{t\in\mathcal T}$ using
\eqref{eq:waterfilling}, and evaluate
$F^{0}=\sum_{t=1}^{T}R_t(\mathbf z_t^{0},P_t^{0})$.\\

\While{$q<I_{\max}$}{
    Evaluate the gradients required in
    \eqref{prob:joint_pa_update} at
    $\{\mathbf z_t^{q},P_t^{q}\}_{t\in\mathcal T}$;\

    Solve \eqref{prob:joint_pa_update} to obtain
    $\{\widehat{\mathbf d}_t^{q}\}_{t\in\mathcal T}$;\

    Determine $\alpha_q$ via backtracking and update
    $\mathbf z_t^{q+1}
    =\mathbf z_t^{q}
    +\alpha_q\widehat{\mathbf d}_t^{q}$,
    $\forall t\in\mathcal T$;\

    Update the transmit powers using
    \eqref{eq:waterfilling} to obtain
    $\{P_t^{q+1}\}_{t\in\mathcal T}$;\

    Evaluate
    $F^{q+1}
    =\sum_{t=1}^{T}
    R_t(\mathbf z_t^{q+1},P_t^{q+1})$;\

    \If{$
    \displaystyle
    \frac{|F^{q+1}-F^{q}|}
    {\max\{1,|F^{q}|\}}
    \leq\epsilon
    $}{
        $q\leftarrow q+1$;\

        \textbf{break};
    }

    $q\leftarrow q+1$;
}

\textbf{Output:}
$\{\mathbf z_t^{\star},P_t^{\star}\}_{t\in\mathcal T}
=
\{\mathbf z_t^{q},P_t^{q}\}_{t\in\mathcal T}$.

\caption{Proposed AO-Based Joint Power and PA-Position Optimization}
\label{alg:AO}
\end{algorithm}

\subsection{Complexity and Convergence Analysis}
\label{subsec:complexity_convergence}

We analyze the computational complexity and convergence of the proposed
AO algorithm. Let $N=N_{\mathrm{Tx}}+N_{\mathrm{Rx}}$ denote the total
number of PAs per time slot. The power-allocation step requires a
one-dimensional bisection search for the dual variable $\nu$, with
complexity $\mathcal{O}\!\left(T\log(1/\epsilon_{\nu})\right)$, where
$\epsilon_{\nu}$ is the bisection accuracy. The PA-position subproblem
in \eqref{prob:joint_pa_update} is a convex QP with $N$ variables per
time slot. Since the $T$ position updates are separable, the computational
cost of a generic interior-point implementation scales approximately as
$\mathcal{O}\!\left(TI_{\mathrm{IP}}N^3\right)$ per AO iteration,
where $I_{\mathrm{IP}}$ denotes the number of interior-point iterations.
Thus, the overall computational cost scales approximately as
$
\mathcal{O}\!\left(
I_{\mathrm{AO}}T
\left[
\log(1/\epsilon_{\nu})+I_{\mathrm{IP}}N^3
\right]
\right).
$

The convergence of the objective sequence follows from the monotonicity
of the AO updates. The backtracking procedure accepts only feasible
PA-position updates that do not decrease the communication objective,
while the subsequent power-allocation step optimally solves the
corresponding power subproblem. Hence,
$
F(\mathbf z^{q+1},\mathbf P^{q+1})
\geq
F(\mathbf z^{q+1},\mathbf P^{q})
\geq
F(\mathbf z^{q},\mathbf P^{q}),
$
so the objective sequence generated by Algorithm~\ref{alg:AO} is
nondecreasing. Moreover, the feasible powers and PA positions are
bounded by the constraints of Problem~\eqref{prob:main}, and the
communication objective is continuous over the resulting bounded
feasible set. Therefore, the objective sequence is upper bounded and
converges to a finite value.

\section{Numerical Results}
\label{sec:numerical_results}

In this section, numerical results are presented to evaluate the
performance of the proposed dual-waveguide PA-assisted ISAC framework.
The results characterize the communication--localization trade-off and
demonstrate the benefits of jointly optimizing the transmit power and
the transmit and receive PA positions.

\subsection{Simulation Setup}

Following commonly adopted settings in the PA literature
\cite{11481959,11122551,11111111, 11212813, ref_conference, yang2026noma}, we consider a carrier frequency of
$f_c=28~\mathrm{GHz}$, an effective refractive index of
$n_{\mathrm{eff}}=1.4$, and a waveguide height of $3~\mathrm{m}$.
The transmit and receive waveguides have equal lengths of $10~\mathrm{m}$
and are located at $y_{\mathrm{Tx}}=-1~\mathrm{m}$ and
$y_{\mathrm{Rx}}=1~\mathrm{m}$, respectively. Each waveguide is
equipped with six PAs with minimum adjacent-PA spacing
$\delta_i=\lambda_g/2$. Equal coupling is assumed on each waveguide,
i.e., $\rho_k^{(i)}=1/N_i$, and the in-waveguide attenuation
coefficient is set to $\kappa_i=0.0092~\mathrm{m}^{-1}$.
The user/target location is uniformly distributed over the rectangular
service region $\mathcal{A}=[-5,5]\times[8,15]~\mathrm{m}^2$. Unless
otherwise specified, the reported results are averaged over 500
independent Monte Carlo realizations of the user/target location. The effective sensing coefficient is normalized as
$|\beta_t|=1$.
The main simulation parameters are summarized in
Table~\ref{tab:simulation_parameters}.

\begin{table}[t]
\centering
\caption{Simulation Parameters}
\label{tab:simulation_parameters}
\renewcommand{\arraystretch}{1.12}
\setlength{\tabcolsep}{4.5pt}
\begin{tabular}{lc}
\toprule
\textbf{Parameter} & \textbf{Value} \\
\midrule

Carrier frequency $f_c$
& $28~\mathrm{GHz}$ \\

Bandwidth $B$
& $100~\mathrm{MHz}$ \\

Effective refractive index $n_{\mathrm{eff}}$
& $1.4$ \\

Waveguide lengths $(L_{\mathrm{Tx}},L_{\mathrm{Rx}})$
& $(10,10)~\mathrm{m}$ \\

Waveguide heights $(d_{\mathrm{Tx}},d_{\mathrm{Rx}})$
& $(3,3)~\mathrm{m}$ \\

Lateral coordinates $(y_{\mathrm{Tx}},y_{\mathrm{Rx}})$
& $(-1,1)~\mathrm{m}$ \\

Number of TPAs/RPAs
& $6/6$ \\

Minimum PA spacing $\delta_i$
& $\lambda_g/2$ \\

PA coupling coefficient $\rho_k^{(i)}$
& $1/N_i$ \\

In-waveguide attenuation $\kappa_i$
& $0.0092~\mathrm{m}^{-1}$ \\

Service region $\mathcal{A}$
& $[-5,5]\times[8,15]~\mathrm{m}^2$ \\

Maximum transmit power $P_{\max}$
& $30~\mathrm{dBm}$ \\

Noise powers $(\sigma_c^2,\sigma_s^2)$
& $(-90,-90)~\mathrm{dBm}$ \\

Total energy budget $E_{\max}$
& $10~\mathrm{J}$ \\

Number of time slots $T$
& $20$ \\

Slot duration $\Delta t$
& $1~\mathrm{s}$ \\

Number of realizations
& $500$ \\

Sensing-SNR threshold $\Gamma_{\min}$
& $10~\mathrm{dB}$ \\

CRB threshold $\varepsilon_{\max}$
& $10^{-2}~\mathrm{m}^2$ \\

\bottomrule
\end{tabular}
\end{table}

\subsection{Benchmark Schemes}

To evaluate the proposed design, the following benchmark schemes are
considered.

\begin{itemize}
    \item \textbf{Single-Waveguide:} The transmit and receive PAs share
    a common waveguide with identical longitudinal positions, i.e.,
    $\mathbf{x}_t^{(\mathrm{Tx})}=\mathbf{x}_t^{(\mathrm{Rx})}$,
    while the PA positions and transmit power are jointly optimized.

    \item \textbf{Tx-Only Optimization:} The transmit power and TPA
    positions are jointly optimized, while the RPAs are uniformly
    deployed and fixed.

    \item \textbf{Rx-Only Optimization:} The transmit power and RPA
    positions are jointly optimized, while the TPAs are uniformly
    deployed and fixed.

    \item \textbf{Fixed-PA:} The TPAs and RPAs are uniformly deployed
    and fixed, while only the transmit power is optimized.

\item \textbf{Equal-Power Allocation:} The transmit and receive PA
positions are jointly optimized, while the available transmit energy
is equally allocated across the time slots.
    
\end{itemize}

\begin{figure}[t]
\centering
\includegraphics[width=7.25cm,height=4.75cm]{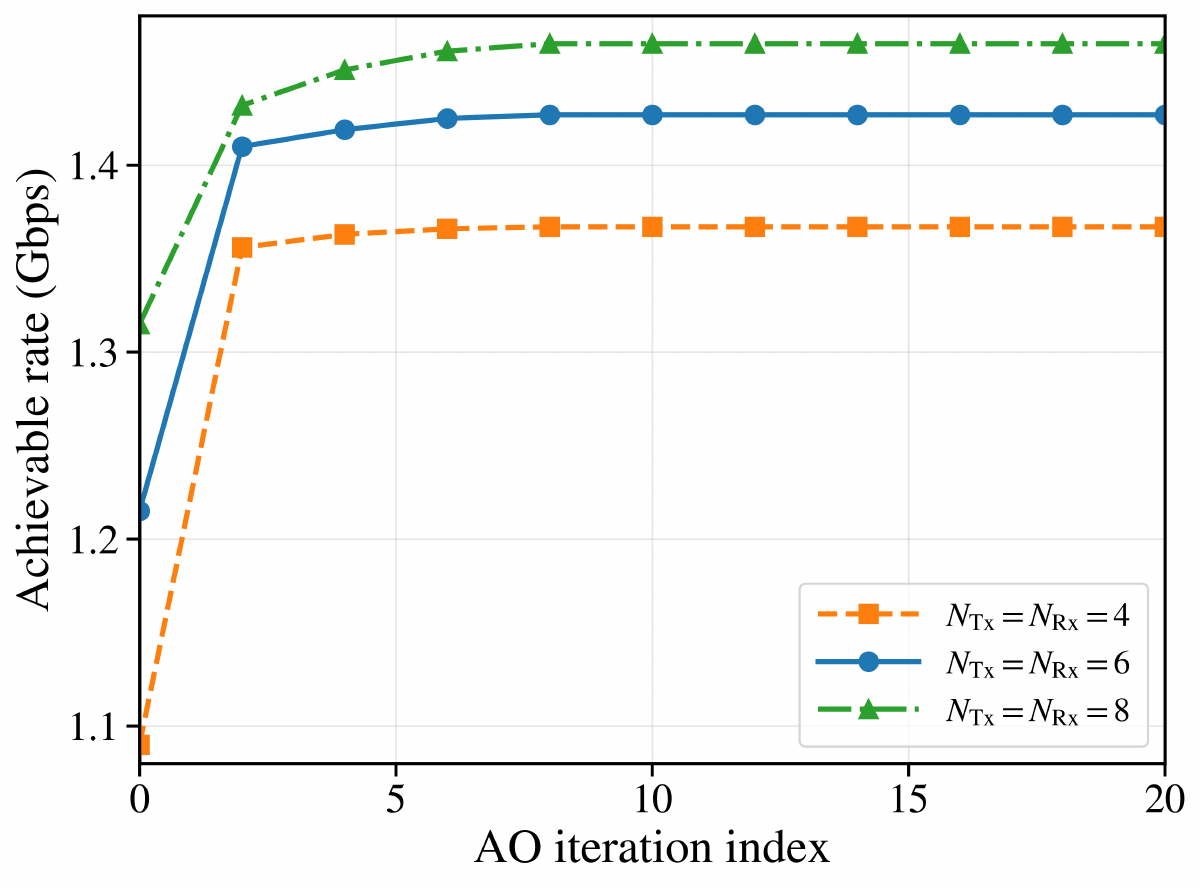}
\caption{Convergence behavior of the proposed algorithm.}
\label{fig:convergence}
\end{figure}

Fig.~\ref{fig:convergence} illustrates the convergence behavior of the
proposed AO algorithm for different numbers of transmit and receive PAs. For all considered configurations, the achievable rate increases
monotonically over the AO iterations and approaches a stable value. Moreover, increasing the number of PAs improves the converged achievable rate, reflecting the additional spatial configurability provided by larger PA deployments.

\begin{figure}[t]
\centering
\includegraphics[width=7.25cm,height=4.75cm]{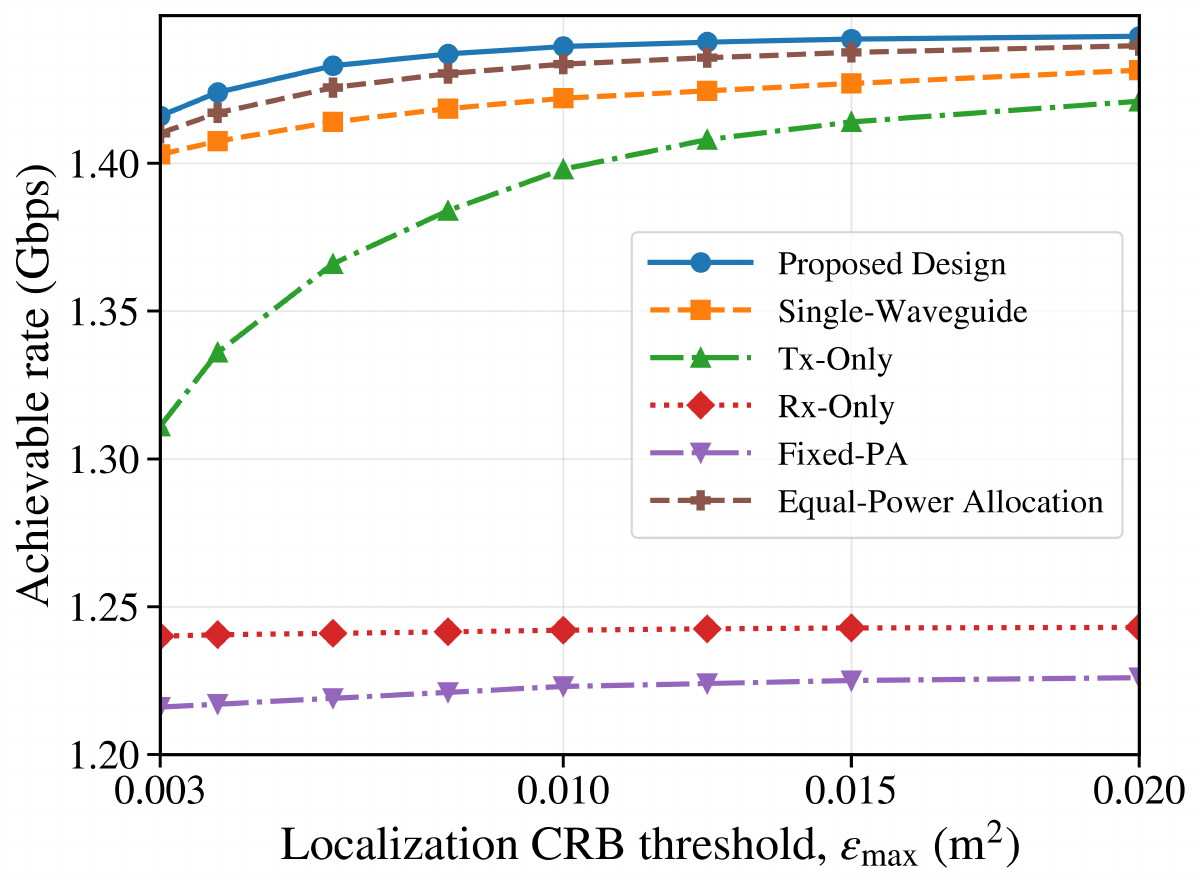}
\caption{Achievable rate versus the localization CRB threshold.}
\label{fig:rate_vs_crb}
\end{figure}

Fig.~\ref{fig:rate_vs_crb} shows the achievable rate versus the
localization CRB threshold, $\varepsilon_{\max}$. As the CRB requirement
is relaxed, the achievable rate increases and gradually saturates. The
proposed design consistently achieves the highest rate through joint
optimization of the transmit and receive PA positions and transmit
power, with a more pronounced gain under stringent localization
requirements. The Equal-Power Allocation benchmark performs below the
proposed design, demonstrating the benefit of adaptive power allocation
across time slots. The Tx-Only scheme is more sensitive to
$\varepsilon_{\max}$ due to its fixed receive configuration, whereas
the Single-Waveguide scheme approaches the proposed design as the CRB
constraint is relaxed. In contrast, the Rx-Only and Fixed-PA schemes
exhibit only marginal improvements with increasing
$\varepsilon_{\max}$ due to their limited communication-side spatial
flexibility.

\begin{figure}[t]
\centering
\includegraphics[width=7.25cm,height=4.75cm]{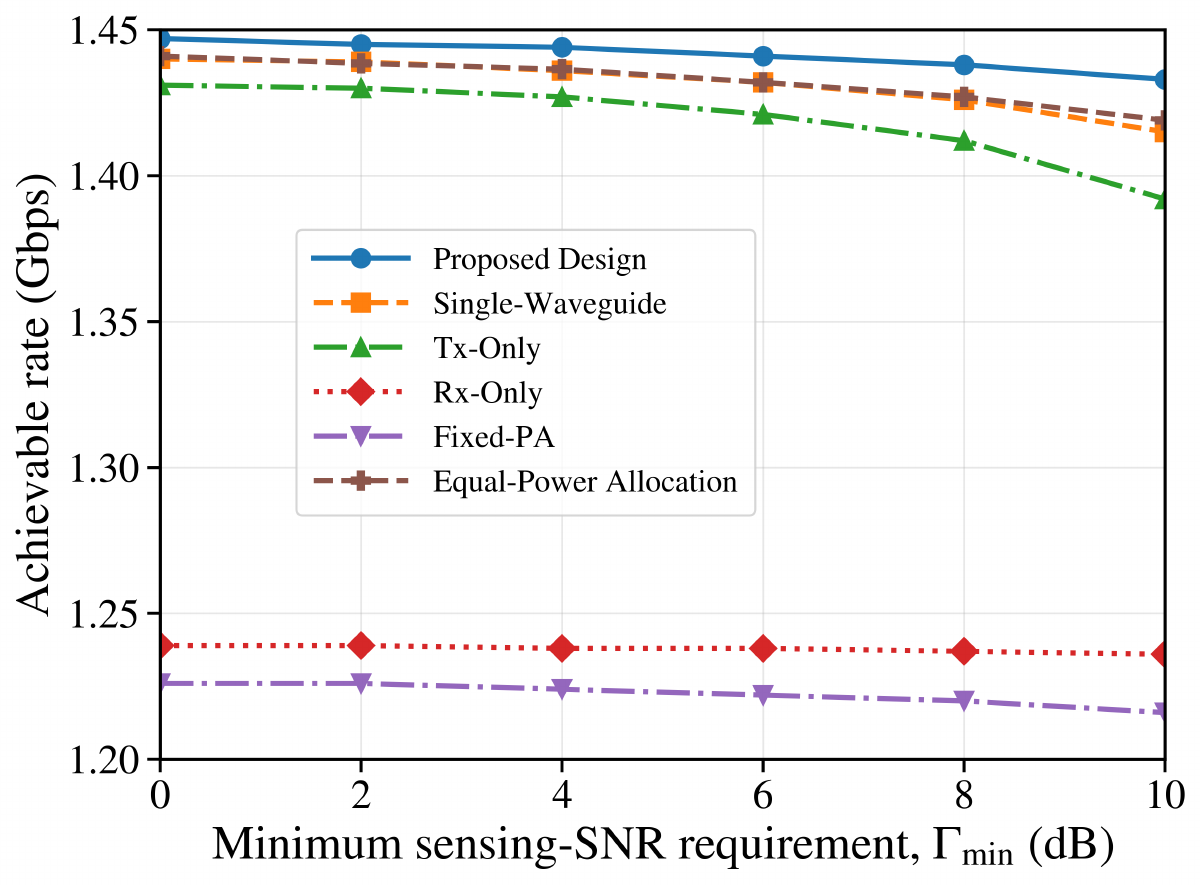}
\caption{Achievable rate versus the minimum sensing-SNR requirement.}
\label{fig:rate_vs_sensing_snr}
\end{figure}

Fig.~\ref{fig:rate_vs_sensing_snr} shows the achievable rate versus the
minimum sensing-SNR requirement, $\Gamma_{\min}$. As $\Gamma_{\min}$
increases, the sensing constraint becomes more stringent, reducing the
flexibility available for communication and hence the achievable rate.
The proposed design consistently achieves the highest rate by jointly
optimizing the transmit and receive PA positions and transmit power.
The Equal-Power Allocation benchmark exhibits an increasing performance
gap as the sensing requirement becomes more stringent, highlighting the
benefit of adaptive power allocation in balancing communication and
sensing requirements. The Single-Waveguide and Tx-Only schemes incur
larger rate losses due to their reduced spatial flexibility in
configuring the sensing paths, whereas the Rx-Only and Fixed-PA schemes
remain substantially limited by their fixed transmit PA configurations.

\begin{figure}[t]
\centering
\includegraphics[width=7.25cm,height=4.75cm]{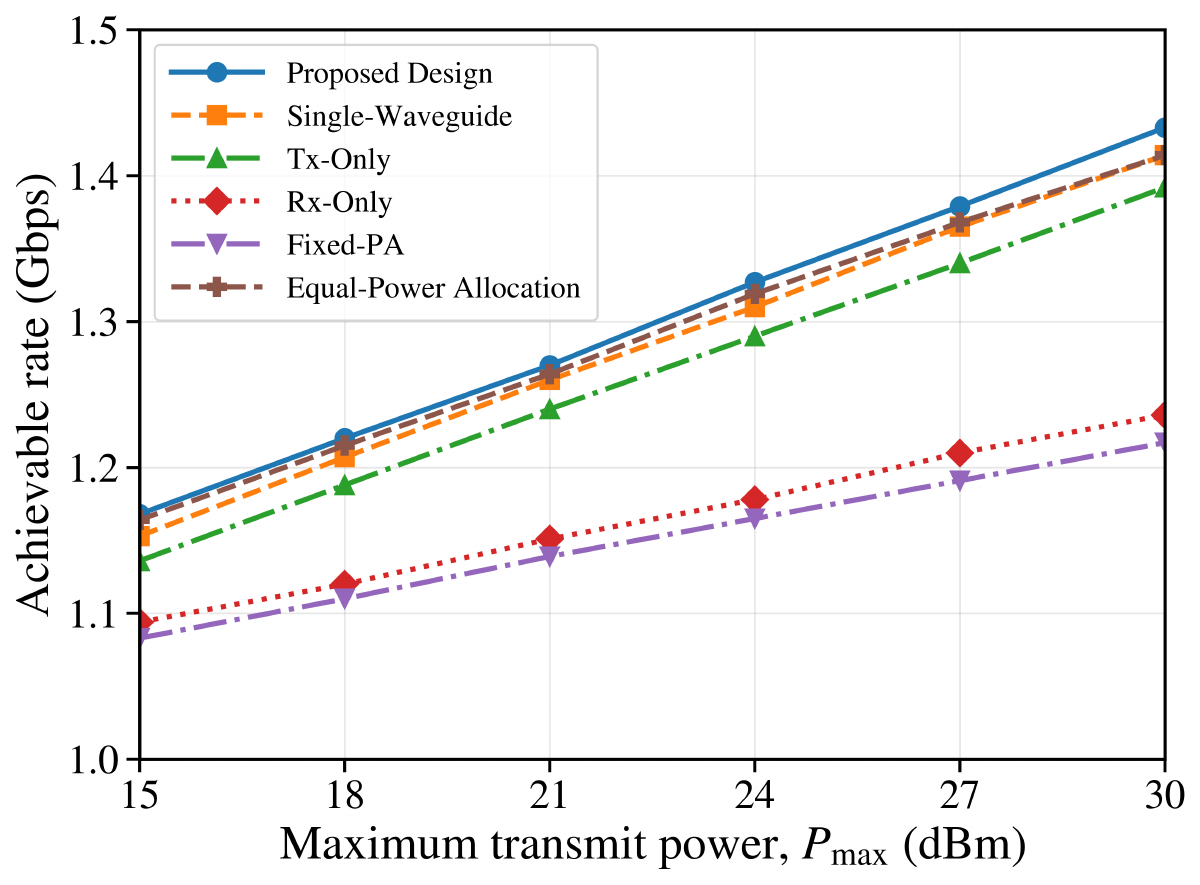}
\caption{Achievable rate versus the maximum transmit power.}
\label{fig:rate_vs_pmax}
\end{figure}

Fig.~\ref{fig:rate_vs_pmax} shows the achievable rate versus the
maximum transmit power, $P_{\max}$. As $P_{\max}$ increases, the
per-slot power constraint is relaxed, leading to higher achievable
rates for all schemes. The proposed design consistently achieves the
highest rate by jointly optimizing the transmit and receive PA
positions and transmit power. Its increasing advantage over the
Equal-Power Allocation benchmark highlights the benefit of adapting
the transmit power across time slots as the per-slot power budget
becomes less restrictive. The Single-Waveguide and Tx-Only schemes
achieve lower rates due to their reduced spatial flexibility, whereas
the Rx-Only and Fixed-PA schemes exhibit more limited gains because
their transmit PA configurations cannot be jointly adapted.

\begin{figure}[t]
\centering

\includegraphics[width=7.25cm,height=4.75cm]{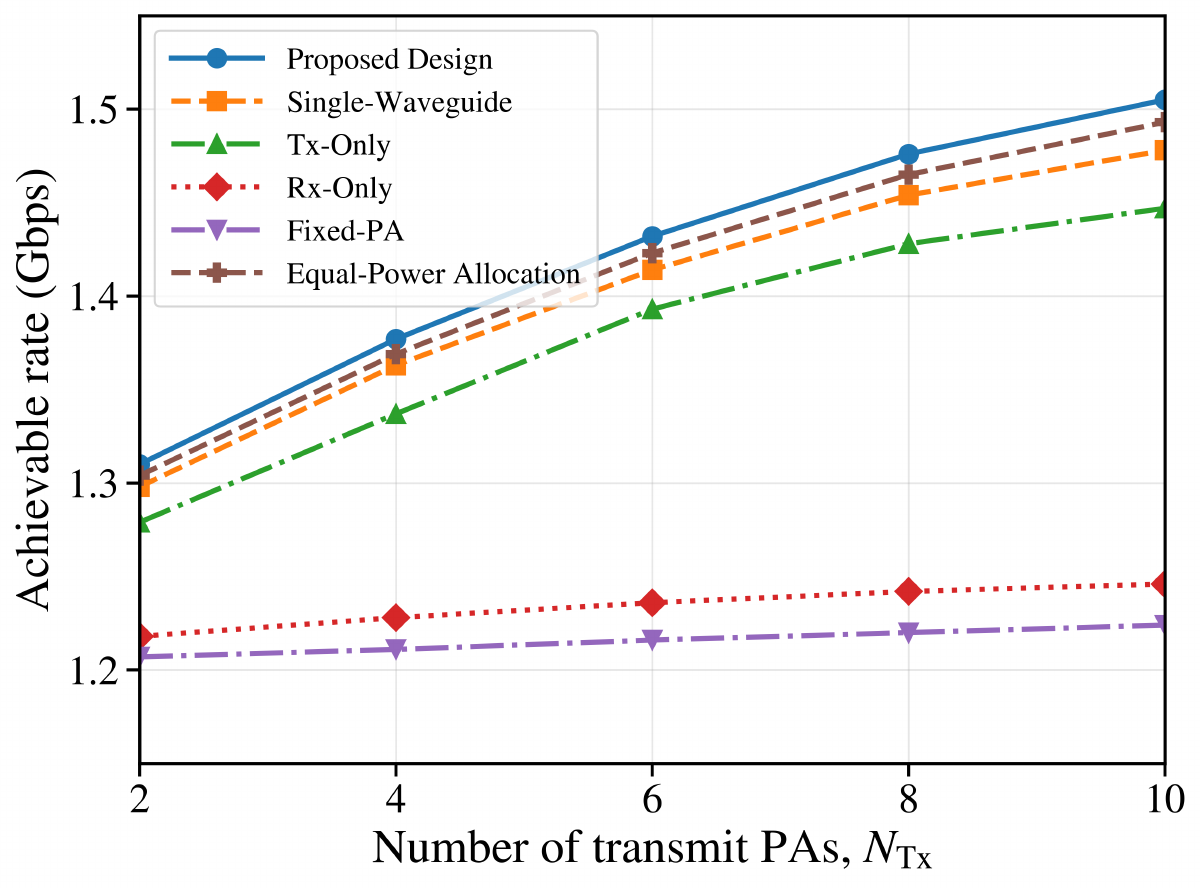}\\[-1mm]
{\small (a)}\\[1mm]

\includegraphics[width=7.25cm,height=4.75cm]{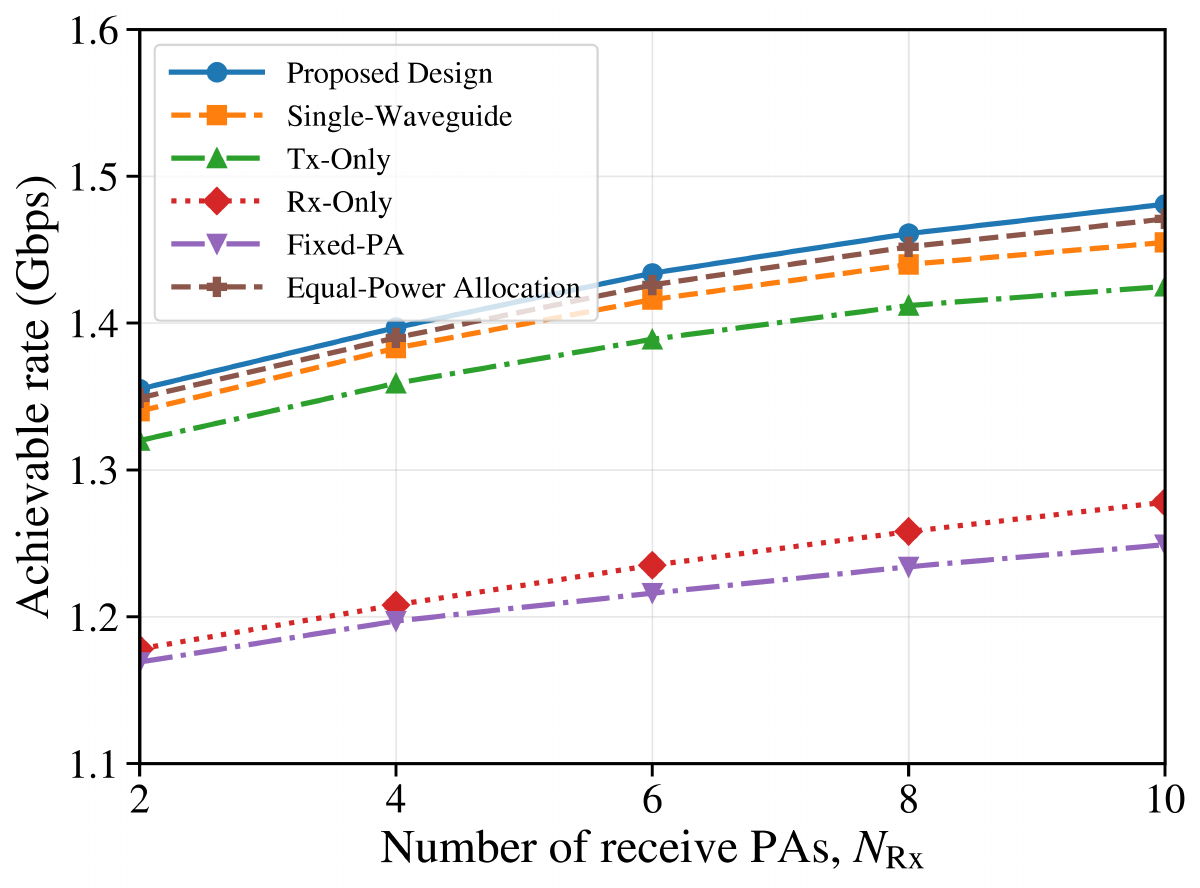}\\[-1mm]
{\small (b)}

\caption{Achievable rate versus the number of (a) transmit PAs for
$N_{\rm Rx}=6$ and (b) receive PAs for $N_{\rm Tx}=6$.}
\label{fig:rate_vs_num_pas}
\end{figure}

Fig.~\ref{fig:rate_vs_num_pas} examines the impact of the numbers of
transmit and receive PAs on the achievable rate. As shown in
Fig.~\ref{fig:rate_vs_num_pas}(a), increasing $N_{\rm Tx}$ yields a
substantial rate improvement by enhancing the transmit-side spatial
configurability under the normalized coupling model, although the
incremental gain diminishes as $N_{\rm Tx}$ grows. In contrast,
Fig.~\ref{fig:rate_vs_num_pas}(b) shows a more moderate improvement with
increasing $N_{\rm Rx}$. This difference reflects the distinct roles of
the two PA sets: the TPAs directly shape the communication channel,
whereas the RPAs influence the achievable rate indirectly through the
sensing and localization constraints. Increasing $N_{\rm Rx}$ therefore
provides greater receive-side sensing flexibility while yielding a
smaller communication-rate gain. In both cases, the proposed design
consistently achieves the highest rate, while the Equal-Power Allocation
benchmark remains slightly inferior, reflecting the additional benefit
of adaptive power allocation as the spatial resources increase.

\section{Conclusion}

This paper investigated a dual-waveguide PA-assisted ISAC system with
joint transmit and receive PA-position and transmit-power optimization. By exploiting
the distinct spatial roles of the transmit and receive PA deployments,
the proposed design provides additional flexibility for balancing
communication performance with sensing and localization requirements.
An AO-based solution was developed by combining a bounded water-filling
power update with a proximal first-order joint PA-position update. Numerical results demonstrated consistent
rate gains over the considered benchmark schemes, particularly under
stringent sensing and localization requirements. They further revealed
the distinct impact of transmit- and receive-side spatial resources on
system performance. Overall, the results highlight the potential of
independently reconfigurable transmit and receive PA deployments for
enhancing the communication--sensing trade-off in PA-assisted ISAC
systems.

\appendices

\section{}
\label{app:power_derivation}

For fixed transmit- and receive-PA positions, using the definitions of
$a_t$ and $P_t^{\min}$ in \eqref{eq:at_definition} and
\eqref{eq:minimum_feasible_power}, respectively, the power-allocation
subproblem reduces to
\begin{subequations}
\begin{align}
    \max_{\{P_t\}}\quad&
    \sum_{t=1}^{T} B\log_2(1+a_tP_t)
    \label{app:power_obj}\\
    \mathrm{s.t.}\quad&
    P_t^{\min}\leq P_t\leq P_{\max},
    \quad \forall t,
    \label{app:power_bounds}\\
    &
    \sum_{t=1}^{T}P_t\Delta t\leq E_{\max}.
    \label{app:power_energy}
\end{align}
\end{subequations}
Provided that the fixed PA configuration admits a feasible power allocation, the above problem is convex. Let
$\nu\geq0$ denote the dual variable associated with
\eqref{app:power_energy}, and let $\lambda_t^{-},\lambda_t^{+}\geq0$
correspond to the lower and upper power bounds. The KKT stationarity
condition is
\begin{equation}
    \frac{Ba_t}{\ln 2\,(1+a_tP_t)}
    -\nu\Delta_t+\lambda_t^{-}-\lambda_t^{+}=0.
    \label{app:kkt_stationarity}
\end{equation}
For an interior solution,
$P_t^{\min}<P_t<P_{\max}$, complementary slackness gives
$\lambda_t^{-}=\lambda_t^{+}=0$, and hence
\begin{equation}
    P_t=
    \frac{B}{\nu\Delta t\ln 2}
    -\frac{1}{a_t}.
\end{equation}
Applying the lower and upper bounds yields
\begin{equation}
    P_t^\star(\nu)
    =
    \min\left\{
        P_{\max},
        \max\left[
            P_t^{\min},
            \frac{B}{\nu\Delta t\ln 2}
            -\frac{1}{a_t}
        \right]
    \right\},
    \label{app:bounded_waterfilling}
\end{equation}
which gives \eqref{eq:waterfilling}.

When the total-energy constraint is active, $\nu>0$ is selected such that
$   \sum_{t=1}^{T}P_t^\star(\nu)\Delta t=E_{\max}.
$
Since the allocated energy is nonincreasing in $\nu$, the corresponding
dual variable can be efficiently obtained via bisection. If
$\sum_{t=1}^{T}P_{\max}\Delta t\leq E_{\max}$, the total-energy constraint is inactive and, since the objective is
increasing in each $P_t$, the optimal solution is
$P_t^\star=P_{\max}$ for all $t\in\mathcal{T}$.

\section{}
\label{app:position_derivatives}

This appendix provides the analytical derivatives required for the
joint PA-position update in \eqref{prob:joint_pa_update}. For compactness,
consider the channel coefficient associated with the $k$-th PA of
waveguide $i\in\{\mathrm{Tx},\mathrm{Rx}\}$,
\begin{equation}
a_{k,t}^{(i)}
=
\sqrt{\rho_k^{(i)}}\frac{\alpha}{r_{k,t}^{(i)}}
e^{-\kappa_i\ell_{k,t}^{(i)}}
e^{-j\phi_{k,t}^{(i)}},
\label{eq:app_channel}
\end{equation}
where
\begin{equation}
\phi_{k,t}^{(i)}
=
\frac{2\pi}{\lambda}r_{k,t}^{(i)}
+
\frac{2\pi}{\lambda_g}\ell_{k,t}^{(i)}.
\label{eq:app_phase}
\end{equation}
Using
$\partial r_{k,t}^{(i)}/\partial x_{k,t}^{(i)}
=(x_{k,t}^{(i)}-x_t)/r_{k,t}^{(i)}$
and
$\partial\ell_{k,t}^{(i)}/\partial x_{k,t}^{(i)}=1$,
differentiation of \eqref{eq:app_channel} yields
\begin{equation}
\frac{\partial a_{k,t}^{(i)}}{\partial x_{k,t}^{(i)}}
=
-a_{k,t}^{(i)}
\left[
\left(
\frac{1}{r_{k,t}^{(i)}}+
j\frac{2\pi}{\lambda}
\right)
\frac{x_{k,t}^{(i)}-x_t}{r_{k,t}^{(i)}}
+
\kappa_i
+
j\frac{2\pi}{\lambda_g}
\right].
\label{eq:app_channel_derivative}
\end{equation}
The corresponding derivative of the effective channel is
$\partial g_t^{(i)}/\partial x_{k,t}^{(i)}
=
\partial a_{k,t}^{(i)}/\partial x_{k,t}^{(i)}$.

For any real PA coordinate $z$, the communication-rate derivative is
\begin{equation}
\frac{\partial R_t}{\partial z}
=
\frac{B P_t}
{\ln 2\left(\sigma_c^2+
P_t|g_t^{(\mathrm{Tx})}|^2\right)}
\frac{\partial |g_t^{(\mathrm{Tx})}|^2}{\partial z},
\label{eq:app_rate_derivative}
\end{equation}
where
\begin{equation}
\frac{\partial |g_t^{(i)}|^2}{\partial z}
=
2\Re\left\{
\big(g_t^{(i)}\big)^*
\frac{\partial g_t^{(i)}}{\partial z}
\right\}.
\label{eq:app_channel_power_derivative}
\end{equation}
Accordingly, \eqref{eq:app_rate_derivative} is zero for a receive-PA
coordinate.

Using the sensing-SNR expression in \eqref{eq:sensing_snr_dual}, its
derivative with respect to an arbitrary PA coordinate $z$ is
\begin{equation}
\frac{\partial\gamma_t^{\mathrm{s}}}{\partial z}
=
\frac{P_t|\beta_t|^2}{\sigma_s^2}
\frac{\partial}{\partial z}
\left(
|g_t^{(\mathrm{Tx})}|^2
|g_t^{(\mathrm{Rx})}|^2
\right).
\label{eq:app_snr_derivative}
\end{equation}
Hence, for $i\in\{\mathrm{Tx},\mathrm{Rx}\}$,
\begin{equation}
\frac{\partial\gamma_t^{\mathrm{s}}}
{\partial x_{k,t}^{(i)}}
=
\frac{2P_t|\beta_t|^2}{\sigma_s^2}
|g_t^{(\bar i)}|^2
\Re\left\{
\big(g_t^{(i)}\big)^*
\frac{\partial g_t^{(i)}}
{\partial x_{k,t}^{(i)}}
\right\},
\label{eq:app_snr_pa_derivative}
\end{equation}
where $\bar i$ denotes the waveguide other than $i$.

To evaluate the CRB gradient, recall that
$\boldsymbol{\xi}_t=[x_t,y_t]^{\mathsf T}$ and define
$
\mathbf{v}_t
\triangleq
\frac{\partial \mu_t}{\partial \boldsymbol{\xi}_t}.
$
The FIM can then be expressed compactly as
$
\mathbf{J}_t
=
\frac{2}{\sigma_s^2}
\operatorname{Re}
\left\{
\mathbf{v}_t^*
\mathbf{v}_t^{\mathsf T}
\right\}.
$
For any PA coordinate $z$,

\begin{equation}
\frac{\partial \mathbf{J}_t}{\partial z}
=
\frac{2}{\sigma_s^2}
\operatorname{Re}
\left\{
\left(
\frac{\partial \mathbf{v}_t}{\partial z}
\right)^*
\mathbf{v}_t^{\mathsf T}
+
\mathbf{v}_t^*
\left(
\frac{\partial \mathbf{v}_t}{\partial z}
\right)^{\mathsf T}
\right\},
\label{eq:app_fim_derivative}
\end{equation}
where
$
\frac{\partial \mathbf{v}_t}{\partial z}
=
\frac{\partial}{\partial \boldsymbol{\xi}_t}
\left(
\frac{\partial \mu_t}{\partial z}
\right).
$
Finally, for $\mathbf J_t\succ 0$, differentiating
$\mathrm{CRB}_t=\mathrm{tr}(\mathbf J_t^{-1})$ yields
\begin{equation}
\frac{\partial\mathrm{CRB}_t}{\partial z}
=
-\mathrm{tr}
\left(
\mathbf J_t^{-1}
\frac{\partial\mathbf J_t}{\partial z}
\mathbf J_t^{-1}
\right).
\label{eq:app_crb_derivative}
\end{equation}
These expressions provide the analytical derivatives required for the
joint PA-position update.

\bibliographystyle{IEEEtran}
\bibliography{Main_Document}

\end{document}